\documentclass[aps,prl,twocolumn,superscriptaddress,showpacs]{revtex4}

\usepackage{graphicx}

\begin{document}

%\preprint{}

\title{Direct Observation of Tunneling and Nonlinear Self-Trapping in a single Bosonic Josephson Junction}


\author{Michael Albiez}
\author{Rudolf Gati}
\author{Jonas F\"olling}
\author{Stefan Hunsmann}
\affiliation{Kirchhoff-Institut f\"ur Physik, Universit\"at
Heidelberg, Im Neuenheimer Feld 227, D-69120 Heidelberg, Germany}
\author{Matteo Cristiani}
\affiliation{INFM, Dipartimento di Fisica
E. Fermi, Largo Pontecorvo 3, I-56127 Pisa, Italy}
\author{Markus K. Oberthaler}
\affiliation{Kirchhoff-Institut f\"ur Physik, Universit\"at
Heidelberg, Im Neuenheimer Feld 227, D-69120 Heidelberg, Germany}

%\thanks{}
%\affiliation{}


\date{\today}

\begin{abstract}
We report on the first realization of a single bosonic Josephson
junction, implemented by two weakly linked Bose-Einstein
condensates in a double-well potential. In order to fully
investigate the nonlinear tunneling dynamics we measure the
density distribution in situ and deduce the evolution of the
relative phase between the two condensates from interference
fringes. Our results verify the predicted nonlinear generalization
of tunneling oscillations in superconducting and superfluid
Josephson junctions. Additionally we confirm a novel nonlinear
effect known as macroscopic quantum self-trapping, which leads to
the inhibition of large amplitude tunneling oscillations.
\end{abstract}

%\keywords{}

\maketitle


Tunneling through a barrier is a paradigm of quantum mechanics and
usually takes place on a nanoscopic scale. A well known phenomenon
based on tunneling is the Josephson effect \cite{Jos:62} between
two macroscopic phase coherent wave functions. This effect has
been observed in different systems such as two superconductors
separated by a thin insulator \cite{Lik:79} and two reservoirs of
superfluid Helium connected by nanoscopic apertures
\cite{Per:97,Suk:01}. In this letter we report on the first
successful implementation of a bosonic Josephson junction
consisting of two weakly coupled Bose-Einstein condensates in a
macroscopic double-well potential.

In contrast to all hitherto realized Josephson junctions in
superconductors and superfluids, in this new system the
interaction between the tunneling particles plays a crucial role.
This nonlinearity gives rise to new dynamical regimes. Anharmonic
Josephson oscillations are predicted \cite{Jan:86,Jack:96,Zap:96},
if the initial population imbalance of the two wells is below a
critical value. The dynamics changes drastically for initial
population differences above the threshold of macroscopic quantum
self-trapping \cite{Mil:97,Sme:97,Rag:99} where large amplitude
Josephson oscillations are inhibited. The two different dynamical
regimes have been experimentally investigated in the context of
Josephson junction arrays \cite{Cat:01,Cat:03,Ank:04}. However,
the small periodicity of the optical lattice does not allow to
resolve individual wells and thus the dynamics between neighboring
sites. Our experimental implementation of a single weak link makes
it possible for the first time to directly observe the density
distribution of the tunneling particles in situ. Furthermore we
measure the evolution of the relative quantum mechanical phase
between both condensates by means of interference \cite{And:97}.

\begin{figure}[h!]
\includegraphics[totalheight=7.5cm]{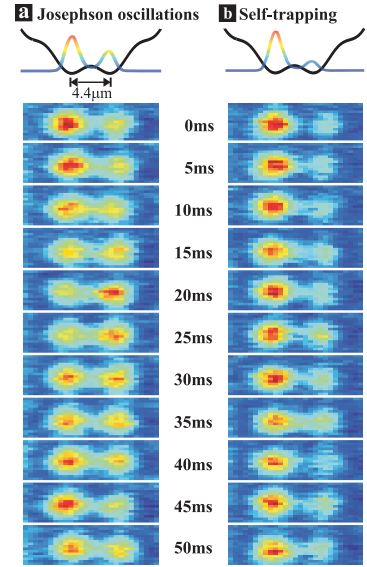}
\caption{Observation of the tunneling dynamics of two weakly
linked Bose-Einstein condensates in a symmetric double-well
potential as indicated in the schematics. The time evolution of
the population of the left and right potential well is directly
visible in the absorption images ($19.4\,\mathrm{\mu m} \times
10.2\,\mathrm{\mu m}$). The distance between the two wavepackets
is increased to $6.7 \mathrm{\mu m}$ for imaging (see text). (a)
Josephson oscillations are observed when the initial population
difference is chosen to be below the critical value $z_C$. (b) In
the case of an initial population difference greater than the
critical value the population in the potential minima is nearly
stationary. This phenomenon is known as macroscopic quantum
self-trapping. \label{figure1}}
\end{figure}

The experimentally observed time evolution of the atomic density
distribution in a symmetric bosonic Josephson junction is shown in
Fig.~\ref{figure1} for two different initial population imbalances
(depicted in the top graphs). In Fig.~\ref{figure1}(a) the initial
population difference between the two wells is chosen to be well
below the self-trapping threshold. Clearly nonlinear Josephson
oscillations are observed i.e. the atoms tunnel right and left
over time. The period of the observed oscillation is 40(2)ms which
is much shorter than the tunneling period of approximately $500$ms
expected for non-interacting atoms in the realized potential. This
reveals the important role of the atom-atom interaction in
Josephson junction experiments with Bose-Einstein condensates. A
different manifestation of the nonlinearity is shown in
Fig.~\ref{figure1}(b) exhibiting macroscopic quantum
self-trapping, which implies that the population imbalance does
not change over time within the experimental errors. The only
difference to the experiment shown in Fig.~\ref{figure1}(a) is
that the initial population imbalance is above the critical value.

The experimental setup and procedure to create the $^{87}$Rb
Bose-Einstein condensates is similar to that used in our previous
work \cite{Eie:04}. A sufficiently precooled thermal cloud is
loaded into an optical dipole trap consisting of two crossed,
focussed laser beams and is subsequently evaporatively cooled by
lowering the light intensities. We produce pure condensates
consisting of $1150 \pm 150$ atoms and final trap frequencies of
$\omega_x = 2\pi\times 78(1)$Hz, $\omega_y=2\pi \times 66(1)$Hz
and $\omega_z=2\pi\times 90(1)$Hz, with gravity acting in the
y-direction. Subsequently we adiabatically ramp up a periodic
one-dimensional light shift potential in x-direction to a depth of
$2\pi\times 412(20)$Hz with periodicity $ 5.2(2)\mu$m realized by
a pair of laser beams at a wavelength of $811$nm crossing at a
relative angle of $9^\circ$. The superposition of this periodic
potential with the strong harmonic confinement creates an
effective double-well potential in x-direction with a barrier
height of $2\pi\times 263(20)$Hz, which splits the initial
condensate into two parts separated by $4.4(2)\mu$m realizing a
single weak link (see schematics in Fig.~\ref{figure1}).

\begin{figure*}
 \includegraphics[totalheight=8.8cm]{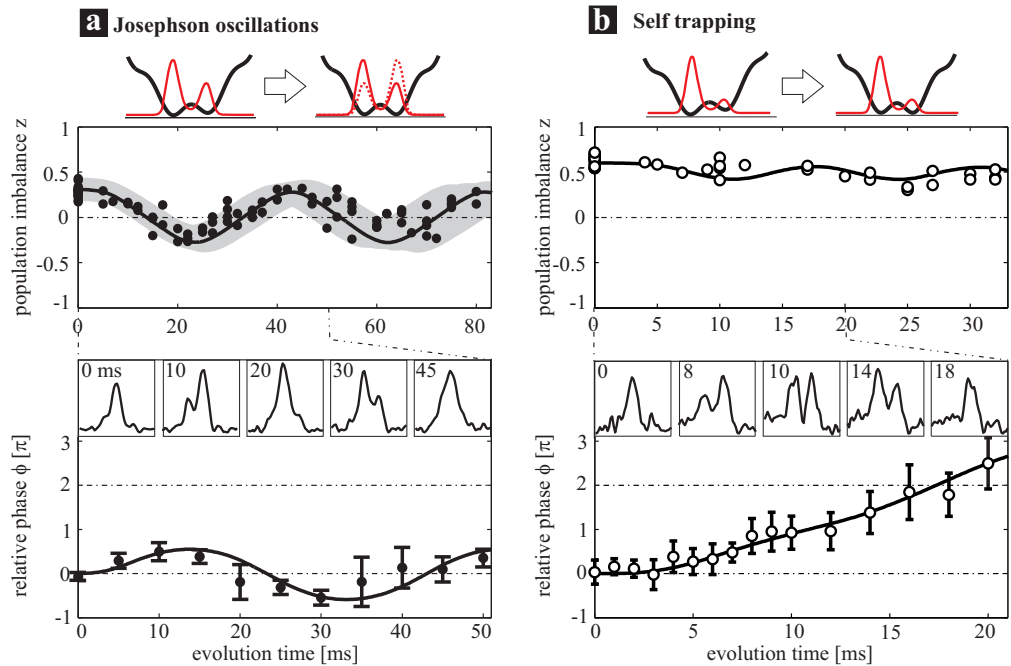}
 \caption{Detailed analysis
of the time dependence of the two dynamical variables $z$ and
$\phi$ describing the system. The top graphs depict the
experimental preparation scheme implemented to realize different
initial atomic distributions. The dynamics is initiated at $t=0$
by switching non-adiabatically to the symmetric double-well
potential. Graph (a) shows the familiar oscillating behavior of
both the population imbalance and the relative phase in the
Josephson regime. The solid lines represent the results obtained
by numerically integrating the non-polynomial Schr\"odinger
equation, and are in excellent agreement with our experimental
findings. The shaded region shows the theoretically expected
scattering of the data due to the uncertainties of the initial
parameters and broadens  for large evolution times due to
different oscillation frequencies for different initial population
imbalances.  The insets depict representative atomic interference
patterns obtained by integrating the absorption images along the
y- and z-direction after the indicated evolution times. In graph
(b) the totally different dynamics in the regime of macroscopic
quantum self-trapping becomes obvious. The population imbalance
exhibits no dynamics within the experimental errors and reveals
the expected nonzero average $\langle z \rangle \neq 0$. Clearly
the phase is unbound and winds up over time. The error bars in the
phase measurements denote statistical errors arising from the
uncertainty of the initial population imbalance.\label{figure2}}
 \end{figure*}

The initial population difference between the left and right
component is obtained by loading the Bose-Einstein condensate into
an asymmetric double-well potential, which is created by a
displacement of the harmonic confinement with respect to the
periodic potential. The asymmetry can be adjusted by shifting the
focussed laser beam which realizes the harmonic confinement in
x-direction. This is done by means of a piezo actuated mirror
mount. A relative shift of only $350$nm leads to a relative
population difference corresponding to the self-trapping
threshold. This demands high passive stability of the mechanical
setup and makes it necessary to actively stabilize the phase of
the periodic potential. With our setup we can adjust any initial
population imbalance with a standard deviation of $\Delta z =
0.06$. The Josephson dynamics is initiated at $t=0$ by
non-adiabatically (with respect to the tunneling dynamics)
changing the potential to a symmetric double-well (see schematics
in Fig.~\ref{figure2}). After a variable evolution time the
potential barrier is suddenly ramped up and the harmonic potential
in x-direction is switched off. This results in dipole
oscillations of the atomic clouds around two neighboring minima of
the periodic potential. Thus by imaging at the time of maximum
separation $(1.5$ms$)$ we can observe clearly distinct wave
packets with a distance of $6.7(5)\mu$m. The atomic density is
deduced from absorption images with a spatial resolution of
$2.2(2)\mu$m. In previously reported realizations of Bose-Einstein
condensates in double-well potentials \cite{Tie:03,Shi:04} the
time scale of tunneling dynamics was in the range of thousands of
seconds.  In contrast, our small interwell distance combined with
a negligible thermal atomic fraction allows the realization of
tunneling times on the order of $50$ms, which makes the direct
observation of the nonlinear dynamics in a single bosonic
Josephson junction possible for the first time.

The physics of Josephson junctions is based on the presence of two
weakly coupled macroscopic wave functions separated by a thin
potential barrier. Insight into the dynamics of the system can be
gained by employing a two mode approximation which characterizes
the wave function by only two parameters, the fractional relative
population $z=(N_l - N_r)/(N_l + N_r)$ and the quantum phase
difference $\phi = \phi_r - \phi_l$ between the left ($l$) and
right ($r$) component. In this framework the resulting quantum
dynamics in a symmetric double-well potential is described by two
coupled differential equations
\begin{eqnarray} \label{eq.1}
\dot z & = & - \sqrt{1-z^2}\sin{\phi}\\
\dot \phi & = & \Lambda z +
\frac{z}{\sqrt{1-z^2}}\cos{\phi}\nonumber
\end{eqnarray}
where $\Lambda$ is proportional to the ratio of the on-site
interaction energy and the coupling matrix element given in
\cite{Sme:97}. These equations represent the nonlinear
generalization of the sinusoidal Josephson oscillations occurring
in superconducting junctions. An intuitive understanding of the
rich dynamics of this system can be gained by considering a
descriptive mechanical analog. The equations given above describe
a classical non-rigid pendulum of tilt angle $\phi$, angular
momentum $z$, and a length proportional to $\sqrt{1-z^2}$. In the
following discussion we will only consider the case of vanishing
initial phase difference $\phi(0)=0$. If the initial population
imbalance is below the critical value \cite{Rag:99} $|z(0)|<z_C$
(from our experimental results we deduce $z_C\approx0.5$
corresponding to $\Lambda\approx15$), equ.~\ref{eq.1} describes
oscillations in $z$ and $\phi$. In the limit of $|z(0)|\ll z_C$
this reduces to a harmonically oscillating mathematical pendulum.
In the context of Josephson junctions this behavior is known as
plasma oscillations. A different dynamical phenomenon arises if
the initial population imbalance is above the critical value. This
implies that the difference between the two on-site interaction
energies becomes larger than the tunneling energy splitting
\cite{footnote}. In this case the relative phase rapidly increases
in time leading to a rapidly alternating tunneling current
according to equ.~\ref{eq.1}. This results in a population
imbalance which performs small oscillations around the initial
value (self-trapping, running phase modes \cite{Rag:99}).

In this case the population difference is self-locked to the
initial value and the relative phase is increasing monotonically
(running phase modes \cite{Rag:99}). In the mechanical analogue
this critical imbalance corresponds to an initial angular momentum
sufficiently large that the pendulum reaches the top position and
continues to rotate with a non vanishing angular momentum.

In order to fully characterize the evolution of the system we
measure not only the population imbalance but also the relative
phase of the macroscopic wave functions. This is achieved by
releasing the Bose-Einstein condensates from the double-well
potential after different evolution times. After a time of flight
of 5ms in the Josephson and 8ms in the self-trapping regime the
wave packets interfere unveiling the relative phase in a direct
way since the resulting atomic fringes are similar to a double
slit diffraction pattern.

In Fig.~\ref{figure2} we present the quantitative analysis of our
experimental results. The measured fractional population imbalance
and the relative phase in the regime of Josephson oscillations
$(z(0)=0.28(6)<z_C)$ are shown in Fig.~\ref{figure2}(a). As
expected for a symmetric double-well potential the relative
population oscillates around its mean value $\langle z\rangle =
0$. The relative phase of the two Bose-Einstein condensates
oscillates with a finite amplitude of $\phi=0.5(2) \pi$ around
$\langle \phi\rangle = 0$. The self-trapping regime can be reached
by simply increasing the initial asymmetry of the double-well
potential as indicated in the schematic diagram in
Fig.~\ref{figure2}(b) realizing $z(0)=0.62(6)>z_C$. In this case
theory predicts that $z$ exhibits only small amplitude
oscillations which never cross $z=0$ i.e. $\langle z\rangle \neq
0$. Additionally the relative phase $\phi$ is unbound and is
supposed to wind up in time. In Fig.~\ref{figure2}(b) these
characteristics of macroscopic quantum self-trapping are evident.
The population difference does not change over time within the
experimental errors and the phase increases monotonically. The
initial deviation from the linear time dependence of the phase is
due to the finite response time
 of the piezo actuated mirror.

The experimentally obtained results can be understood
quantitatively by going beyond the two mode model which assumes
stationary wave functions in the individual wells which is only
justified for $N_l + N_r \ll 1000$ atoms \cite{Mil:97}. Therefore
we numerically integrate the non-polynomial Schr\"odinger equation
\cite{Sal:02} using the independently measured trap parameters and
atom numbers. The calculations also include the fact that the
piezo actuated mirror initiating the Josephson dynamics reaches
its final position only after $7$ms. It is remarkable that all
experimental findings are in excellent quantitative agreement with
our numerical simulation without free parameters.

\begin{figure}
 \includegraphics{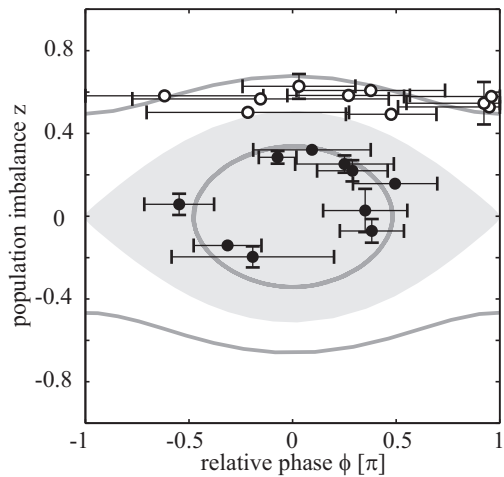}
 \caption{Quantum phase-plane
portrait for the bosonic Josephson junction. In the regime of
Josephson oscillations the experimental data are represented with
filled circles and in the self-trapping regime with open circles.
The shaded region, which indicates the Josephson regime, and the
solid lines are obtained by solving the coupled differential
equ.~\ref{eq.1} with our specific experimental parameters. The two
mode model explains the observed $z(\phi)$ dependence reasonably
in both dynamical regimes. The error bars represent the
statistical error and mainly result from the high sensitivity of
the relative phase on the initial population imbalance especially
for long evolution times.\label{figure3}}
 \end{figure}

The distinction between the two dynamical regimes - Josephson
tunneling and macroscopic self-trapping - becomes very apparent in
the phase-plane portrait of the dynamical variables $z$ and
$\phi$. For our experimental situation this is shown in
Fig.~\ref{figure3} where we compare our results with the
prediction of the simple two mode model.
From our experimental observations the critical population
imbalance can be estimated to $z_C = 0.50(5)$. In the framework of
the two mode model \cite{Rag:99} this yields $\Lambda=15(3)$. The
corresponding solutions of equ.~\ref{eq.1} are depicted with solid
lines. Clearly the basic features of the dynamics are well
captured by this approach. In the nonlinear Josephson tunneling
regime ($z<z_C$) the dynamical variables follow a closed phase
plane trajectory as predicted by the simple model. Very recently a
variable tunneling two mode model has been discussed by Ananikian
and Bergeman \cite{Ana:05} which is quantitative agreement with
our experimental observations.

The successful experimental realization of weakly coupled
Bose-Einstein condensates adds a new tool to quantum optics with
interacting matter waves. It opens up new avenues ranging from the
generation of squeezed atomic states \cite{Dun:01} and entangled
number states (Schr\"odinger cat states) \cite{Mah:03} to
applications such as atom interferometry \cite{And:02}. Moreover
the detailed investigation of the self-trapping phenomenon could
provide a test of the validity of the mean field description in
atomic gases in the strong nonlinear regime \cite{Str:04}.

\begin{acknowledgments}
We wish to thank Andrea Trombettoni, Augusto Smerzi, Tom Bergeman,
and Luis Santos for very stimulating discussions. We would also
like to thank Thomas Anker and Bernd Eiermann for their
contributions to the experimental setup. This work was funded by
Deutsche Forschungsgemeinschaft 'Emmy Noether Programm' and by the
European Union, RTN-Cold Quantum Gases, Contract No.
HPRN-CT-2000-00125. R. G. thanks the Landesgraduiertenf\"orderung
Baden-W\"urttemberg for the financial support.
\end{acknowledgments}

\bibliography{apsbib}

\begin{thebibliography}{24}
\bibitem[{foo()}]{footnote}
\bibinfo{note}{Note: By solving \cite{Sal:02} we obtain a chemical potential in
  the Josephson oscillation regime of $\mu = 2 \pi \times 285 \pm 15$ Hz, which
  is comparable to the potential barrier height}.
\end{thebibliography}

\end{document}